\documentclass[runningheads]{llncs}
\usepackage{notoccite}

\usepackage{graphicx}
\usepackage{multirow}
\usepackage{lscape}
\usepackage[super]{nth}
\usepackage{subfig}
\usepackage{amsmath}
\usepackage{multicol}
\usepackage{listings}
\usepackage{soul}

\usepackage{algorithm}
\usepackage{algorithmic}
\usepackage{multicol}
\usepackage{setspace}

\usepackage{tabularx}
\usepackage{booktabs}
\usepackage{rotating}
\usepackage{forest}

\renewcommand{\arraystretch}{1.5}
\usepackage{amssymb}
\usepackage{pifont}
\usepackage{multirow}

\definecolor{backcolour}{rgb}{0.95,0.95,0.95}
\begin{document}
\title{GLASS: Towards Secure and Decentralized eGovernance Services using IPFS}
%
%
\author{Christos Chrysoulas\inst{1} \orcidID{0000-0001-9817-003X} \and
Amanda Thomson\inst{1} \and
Nikolaos Pitropakis\inst{1}\orcidID{0000-0002-3392-9970} \and
Pavlos Papadopoulos \inst{1}\orcidID{0000-0001-5927-6026} \and 
Owen Lo \inst{1} \and 
William J. Buchanan\inst{1}\orcidID{0000-0003-0809-3523} \and
George Domalis\inst{2} \and
Nikos Karacapilidis\inst{2} \and
Dimitris Tsakalidis\inst{2} \and
Dimitris Tsolis\inst{2}
}
\authorrunning{Christos Chrysoulas et al.}
%
\institute{School of Computing, Edinburgh Napier University,
 Edinburgh, United Kingdom \\
\and
Computer Engineering and Informatics Department, University of Patras, Greece
}
\maketitle              
\vspace{-0.8cm}
\begin{abstract}
The continuously advancing digitization has provided answers to the bureaucratic problems faced by eGovernance services. This innovation led them to an era of automation, broadened the attack surface and made them a popular target for cyber attacks. eGovernance services utilize the internet, which is a location addressed system in which whoever controls its location controls not only the content itself but also the integrity and the access of that content. We propose GLASS, a decentralized solution that combines the InterPlanetary File System with Distributed Ledger Technology and Smart Contracts to secure eGovernance services. We also created a testbed environment where we measure the system's performance.


\end{abstract}

\section{Introduction}
\vspace{-8pt}
The rapid evolution of digital technologies, including mobile communications, cloud computing infrastructures, and distributed applications, has created an extended impact on society while also enabling the establishment of novel eGovernance models. The need for an inclusive eGovernance model with integrated multi-actor governance services is apparent and a key element towards a European Single Market. Digital Transformation of public services can remove existing digital and physical barriers, reduce administrative burdens, enhance governments' productivity, minimize the extra cost of traditional means to increase capacity, and eventually improve the overall quality of interactions with (and within) public administrations.

eGovernance includes novel and digital by default public services aiming for administrative efficiency and minimization of bureaucratic processes, enabling open government capabilities, behavior and professionalism, improved trust and confidence in governmental transactions. Towards the modernization of public services, public administrations need to transform their manual business flows and upgrade their existing internal processes and services.

However, the digitization of eGovernance services has also expanded the attack surface, thus making them attractive to malicious third parties. In 2017 the National Health Service of the United Kingdom suffered from the WannaCry ransomware, which resulted in missed appointments, deaths, and fiscal costs \cite{ghafur2019retrospective}. Recently, in May 2021 the American oil pipeline system suffered a ransomware cyberattack that impacted all the computerized equipment managing the pipeline. The company paid a ransom of 75 Bitcoins, approximately \$5 million, to the hackers in exchange for a decryption tool which eventually proved so slow that Colonial's own backups were used to bring the system back to service \cite{oxford2021efforts}.

As the need for privacy-preserving and secure solutions in eGovernance services is imminent, our decentralized solution, namely GLASS, moves towards that direction by examining the effectiveness and efficiency of distributed cutting edge technologies, demonstrating the capacity of a public, distributed infrastructure, based on the InterPlanetary File System (IPFS). Our contributions can be summarised as follows: 
\vspace{-8pt}
\begin{itemize}
    \item We analyze the threat landscape in the context of an eGovernance use case. 
    \item We create a distributed testbed environment based on IPFS and detail our methodology.
    \item We analyze and critically evaluate the runtime performance of our implementation.
\end{itemize}
\vspace{-8pt}

The structure of the rest of the paper is organized as follows: Section 2 builds the background on distributed models and presents the related literature, while Section 3 details the GLASS architecture while briefly explaining the threat landscape in the context of an eGovernance services use case scenario. Section 4 consists of our methodology and implementation used to conduct the main experimental activity of our work, while Section 5 presents and evaluates the performance results of our experimental activity. Finally, Section 6 draws the conclusions, giving some pointers for future work.
\vspace{-16pt}

\section{Background and Related Literature}
\vspace{-12pt}

\subsection{Kademlia}
\vspace{-8pt}

In 2001 Maymounkov and Mazières published Kademlia, a  Distributed Hash Table (DHT) that offered multiple features that were currently not available simultaneously in any other DHT \cite{xor-metric}. The paper introduced a novel XOR metric to calculate the distance between nodes in the key space and a node Id routing algorithm that enabled nodes to locate other nodes close to a given target key efficiently. The presented single routing algorithm was more optimal compared to other algorithms such as Pastry\cite{pastry}, Tapestry\cite{tapestry} and Plaxton\cite{Plaxton} that all required secondary routing tables. Kademlia was outlined as easily optimised with a base other than 2 with no need for secondary routing tables. The k-bucket table was configured so as to approach the target \(b\) (initial implementation was b = 5 ) bits per hop. With one bucket being used for nodes within distance range of \([j2^{160-(i+1)b)}, (j+1)2^{160-(i+1)b}]\) from the initial node for each \(0 < j < 2^{b}\) and \(0 \leq i < 160/b\) based on a SHA1 160 bit address space. At any point, it is expected that there would be no more than \((2^{b} - 1)log_{2^{b}}\) buckets with entries. The k-buckets were described as being resistant to certain DoS attacks \cite{xor-metric} due to the inability to flood the system with new nodes, as Kademlia only inserts new nodes once old ones leave. 

In 2008 Baumgart and Mies introduced S/Kademlia \cite{S/Kad} which offered several further security enhancements designed to improve on the original specification. They examined various attacks that peer-to-peer networks were vulnerable to and offered practical solutions to protect against them. The key attacks identified by them were:
a) Eclipse Attack, b) Sybil Attack, and c) Adversarial Routing. In 2020 Prünster et al. \cite{eclipse} highlighted the need for further implementation of S/Kademlia mitigations by demonstrating an effective eclipse attack. They were able to generate a large number of ephemeral identities and poison multiple nodes routing tables for very little expense, and \textit{CVE-2020-10937} was assigned to the demonstrated attack.
\vspace{-16pt}

\subsection{IPFS}
\vspace{-8pt}

The InterPlanetary File System (IPFS) is a distributed system based on a peer-to-peer protocol that provides public data storage services to transform the web into a new decentralized and more efficient tool. Its primary purpose is to replace the HTTP protocol for document transactions by solving HTTP's most limiting problems like availability, cost, and centralization of data in data centers. 

IPFS is based on a Merkle Directed Acyclic Graph (DAG) \cite{Kothari_2019}, the data structure to keep track of the location its data chunks are stored and the correlation between them. Each data block has a unique content identifier (CID) fabricated by hashing its content in this peculiar data structure. In case the content of a node's child changes, the CID of the parent node changes as well. For someone to access a file, knowing its unique Content Identifier, constructed by the hash of the data contained within it, is essential. Each participating node (user) keeps a list of the CIDs it hosts in a Distributed Hash Table (DHT) implemented using the Kademlia protocol \cite{10.1007/3-540-45748-8_5}. Each user ``advertises'' the CIDs they store in the DHT, resulting in a distributed ``dictionary'' used for looking up content. When a user tries to access a specific file, IPFS crawls the DHTs to locate the file by matching the unique content identifier. Using content-based addressing instead of location-based addressing serves in preventing saving duplicate files in the network and tracking down a file by its content rather than by its address.

IPFS enables its users to store and distribute data globally in a secure, resilient and efficient way. Each file uploaded on IPFS is fragmented into chunks of 256Kb and hashed before being scattered in participating nodes around the globe. Following the aforementioned methodology, data integrity is ensured since no one can tamper with a data block without affecting its unique hash. Furthermore, data resilience is ensured by placing the same data block in more than one participating node.

Mukne et al. \cite{8944471} are using IPFS and Hyperledger Fabric augmented to perform secure documentation of land record management. Andreev and Daskalov \cite{Andreev2018} are using IPFS to keep students’ personal information off-chain in a solution that manages students’ data through blockchain. Singh \cite{phdthesis2018} created an architecture for open government data where proof-of-concept uses Ethereum for decentralized processing and BigchainDB and IPFS for storage of large volumes of data and files, respectively.
\vspace{-20pt}

\subsection{Distributed ledger}
\vspace{-10pt}

A Distributed Ledger is a distributed database architecture that enables multiple members to maintain their own identical copy of information without the need for validation from a central entity while ensuring data integrity. Transaction data are scattered among multiple nodes using the P2P protocol principles and are synchronized simultaneously in all nodes. 
By providing Identification Management through DLTs, it is ensured that the user has control of their identity records since the information is stored publicly on the ledger instead of the systems of a central authority. Furthermore, since editing information on past transactions on a blockchain system is not supported, protection against unauthorized alteration of the identity records is established. Finally, having a single record of identity information that the user can utilize on multiple occasions minimizes the data duplication on multiple databases \cite{Dunphy2018AFL}.
The second generation of blockchain technologies introduced the smart contracts that act as mini-programs used to automate code deployment when some pre-defined terms are met. 

Our solution, GLASS, combines the advantages of IPFS with those offered by the Distributed Ledgers and Smart Contracts, thus creating a distributed scalable and secure eGovernance infrastructure. Moving towards the first steps of our implementation, we create an IPFS based testbed environment and empirically evaluate its runtime performance.

\vspace{-16pt}

\section{Architecture}
\vspace{-12pt}

\begin{figure}[!h]
    \centering
    \makebox[\textwidth][c]{\includegraphics[width=1.2\textwidth]{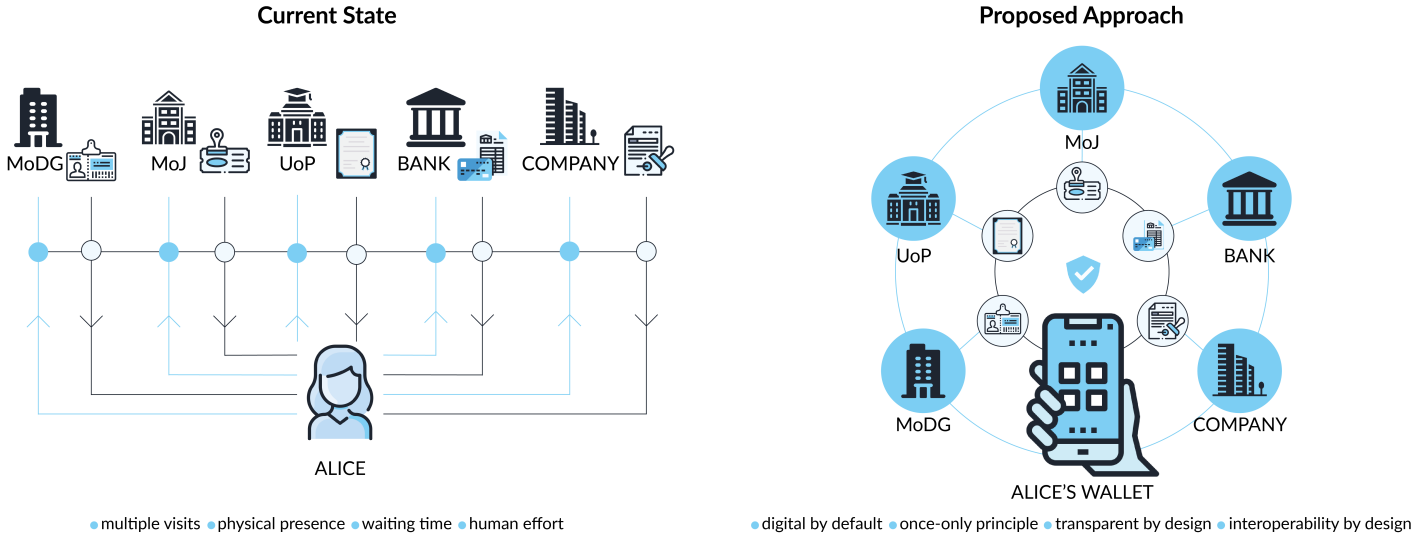}}
    \caption{Current practice compared to our approach \cite{domalis2021trustable}}
    \label{fig:practice_comparison}
\vspace{-5pt}

\end{figure}

We propose a combination of IPFS with Distributed Ledger and Smart Contracts which are proven to be beneficial for recording massive volumes of transactions. Extracting helpful information efficiently has significant computational challenges, such as analysing, aggregating, visualising, and storing data collected in distributed ledgers. More specifically, the volume and velocity of the data make it difficult for typical algorithms to scale while querying the ledger might come at high computation costs. State-of-the-art efforts seek to introduce new models that deal with such large-scale, distributed data queries to reduce data volume transferred over the network via adaptive sampling that maintains certain accuracy guarantees \cite{trihinas2017admin}. As the ledgers (and thus the data) keep getting bigger, a challenge is to make sense of the collected data for the users and perform analytics leveraging big data processing engines (i.e., Spark) that can deliver results quickly and efficiently. In order to adequately protect data resources, it is paramount to encrypt data in such a way that no one other than intended parties should be able to get the original data. The current practice compared to our approach can be seen in Figure~\ref{fig:practice_comparison}.

A simple use case presenting a European Union's citizen, Alice, getting a job abroad from Greece to another member state, Portugal, using GLASS ecosystem, is presented in Algorithm~\ref{alice_portugal}.
\vspace{-16pt}

\setlength{\textfloatsep}{0pt}
\begin{algorithm}[h!]
\caption{Alice getting a job to Portugal}
\begin{spacing}{1.2}
\begin{algorithmic}[1]
\STATE Starting from Greece, Alice finds a vacant job position in Portugal. She applies for the job, and thankfully she gets hired.
\STATE In Portugal, she has to deal with a series of bureaucratic processes (ID card, social security number, open a bank account).
\STATE To obtain a Portuguese Residence title, rent an apartment and open a bank account, Alice needs to present at least a validated ID documentation, birth certificate, nationality certification validated by a Greek Authority and proof that she works in Portugal.
\STATE Adopting the GLASS solution, Alice can request the proof of ID and the validated data from the Ministry of Digital Governance (MoDG).
\STATE The MoDG can issue the document, and after Alice's permission, the document can be forwarded to the Ministry of Justice (MoJ).
\STATE After this transaction is completed, Alice can access and securely share her Portuguese social security number through her Wallet.
\STATE Then Alice's employer in Portugal can directly get the validated social security number from the MoJ, after her approval, to register her credentials to their internal payroll system.
\STATE Using a decentralized application of the GLASS ecosystem, Alice can use her validated digital identity to request remotely the required documentation from the respective Greek Authority (MoDG), the Portuguese authority (MoJ) and her employer.
\STATE MoDG can digitally issue and validate the documentation and transmit the encrypted data into the distributed network while the transaction among the users is being recorded.
\STATE All the transactions, including requests, notifications, and permissions, can be monitored and stored, protecting Alice's (and each participant's) privacy.
\end{algorithmic}
\end{spacing}
\label{alice_portugal}
\end{algorithm}

\subsection{Threat Landscape}
\vspace{-8pt}

Distributed file systems, such as IPFS, need to solve several challenges related to the security and privacy of the stored data, the infrastructure's scalability, the decentralized applications and big data complexities. However, there is a number of promising solutions that aim to settle some of these hurdles.
\vspace{-12pt}

\subsubsection{Security and privacy challenges}
The key challenge of distributed file systems, including IPFS, is that when new peers participate in the system, they can access any stored file, including sensitive documents. Hence, the security and privacy of the system remain an open question, especially due to General Data Protection Regulation (GDPR) \cite{voigt2017eu} in the European Union. A prominent solution to that is the application of smart contract-based Access Control (AC) policies \cite{barati2021design,huang2020bridge,papadopoulos2020privacy,stamatellis2020privacy}, and further encryption mechanisms \cite{wang2018blockchain}.

Another security and privacy challenge is related to file erasure. By their nature, distributed file systems distribute all the stored files and documents among their participating peers. Hence, when data owners transmit ``erasure commands'' to the distributed network, it is not clear if all the peers would obey this command and delete their version of the ``deleted'' file or document. A solution to this data replication issue can be a common technique commonly present in data centers \cite{plank1997tutorial,huang2020blockchain}.
\vspace{-18pt}

\subsubsection{Scalability Challenges}

Since GLASS aims to create an eGovernance framework to be followed by all European Union's member states, the infrastructure's scalability poses a real threat. According to \cite{wennergren2018transparency,shen2019understanding}, one of the scalability issues on IPFS is the bandwidth limit in each IPFS instance due to the peer-to-peer nature of the system. Each participant needs to connect to another IPFS node to read or download the data objects. \cite{nyaletey2019blockipfs} proposed a combination of IPFS and blockchain technology, namely BlockIPFS, to improve the traceability of all the occurred access events on IPFS. The authors measured the latency of each event, such as storing, reading, downloading, by varying the number of IPFS nodes and presented that even incorporating large numbers of IPFS nodes does not significantly improve the latency of all the IPFS actions. However, the authors' experiments were limited to a maximum of 27 nodes; hence, the latency measurement on a vast scale remains an open question.

For the storage optimisation, two prominent solutions can be applied:
\vspace{-7pt}
\begin{itemize}
    \item \textbf{Storing data off-chain} \cite{norvill2018ipfs,poon2017plasma,poon2016bitcoin}. The concept of utilising smart contracts off-chain and use IPFS as a storage database has been presented by some works. This solution is storage efficient since the IPFS nodes need to exchange only hash values of the data.
    \item \textbf{Utilize erasure codes} \cite{rizzo1997effective,wilkinson2014storj,vorick2014sia}. In erasure codes, a file is divided into smaller batches and these batches are encoded. Following that, each batch can be decoded and reconstruct the full file. \cite{chen2017improved} utilized erasure codes in a scenario combining blockchain and IPFS.
\end{itemize}
\vspace{-20pt}

\subsubsection{Decentralized applications complexities}
Multiple novel decentralized applications have already been developed on top of IPFS, with luminous examples, a music streaming platform, and an open-access research publication repository \cite{jia2016opus,tenorio2019towards}. Distributing seemingly centralized applications offer multiple advantages, such as rewarding the creators of music or research publications directly without involving any trusted intermediaries and is feasible with the assistance of blockchain technologies \cite{truong2021blockchain}.

Within the GLASS ecosystem, it is critical to clearly define where these decentralized applications would be developed and executed to avoid obstacles due to the complexities of the underlying technologies. A potential solution is to carry out the execution of the decentralized applications off-chain \cite{truong2021blockchain}, similarly to other popular decentralized applications ecosystems, such as Blockstack \cite{ali2020stacks}.
\vspace{-25pt}

\section{Methodology and Implementation}
\vspace{-14pt}

IPFS uses Libp2p\footnote{Lib2p: https://github.com/libp2p/js-libp2p} as it's base. Originally Libp2p was part of the IPFS project but has since become standalone. It provides all of the transport abstractions and the Kad-DHT functionality. The main release is written in Go, with ports to Rust and JavaScript. To look at the implementation of the DHT, JavaScript was chosen as it natively would not rely on a multi-threading approach but instead asynchronous I/O and an event-driven programming model. 
    
For small scale local testing of the DHT, a simple Libp2p node was created 40 times\footnote{Code can be found at: https://github.com/aaoi990/ipfs-kad-dht-evaluation} to listen on the host, and the port will differentiate each node. The DHT configuration \footnote{DHT configuration: https://github.com/libp2p/js-libp2p-kad-dht} is the standard recommended Libp2p Kad-DHT configuration with all standard defaults applied. The exception being the DHT random walk – which is not enabled by default but does allow for random host discovery. The connection encryption used is Noise protocol. \footnote{Noise Protocol: https://noiseprotocol.org/}

When a new node is initialized, it knows no peers. Typically in IPFS, this issue is solved by bootstrapping the node – providing it with a set of long-serving core nodes that have fully populated routing tables ready to share. In this case, to provide some basic routing entries, the initial node is populated by the address of the next created node, ensuring that each node knows of at least one other but only the next node. Although enabled, the random walk would be an untenable solution to peer discovery in such a small set of nodes given that the Libp2p implementation of the random walk involves dialling a random peerId created from a sha256 multi hash of 16 random bytes.

The last node initialized is then chosen to host the content. To transform the content into a CID, it is first hashed with the standard sha256 algorithm, and then a multi hash is created from this. As we are using CIDv0, the multi hash is then base 58 encoded (CIDv1 is base 32 encoded)and provided to the js-cids library to create the CID.

Once the CID is created, the final node starts providing it to the network. The content routing class of the Kad-DHT will then distribute the pointer to the nodes closest to the key itself. Each node DHT will then begin searching for other nodes and populating its routing table entries. The peer discovery process is best witnessed by examining the debug log for the Kad-DHT by starting the program with the following: \verb|DEBUG="libp2p:dht:*" node index.js|
    
Each instance of the Kad-DHT is initialized with an instance of the Providers class that manages all known providers – a peer known to have the content for a given CID. The providers class is initialized with an instance of the datastore, which houses the records of providers in the format of a key-value pair, with the key being created from the array of the CID and PeerID and the value being the time the record was entered into the store.

When the class is created, it spawns its own cleanup service. The service is a set interval clean up that runs and keeps the list of providers healthy. It is important to note at this point that although a list of providers are stored in the datastore, to ensure access is fast, there is an LRU (least recently used) cache in front of it which speeds up the process of not only cleaning up expired providers but accessing active ones as well. The default constant for the LRU size is 256, and the default cleanup interval is one hour. The cleanup service retrieves all provider entries from the datastore, checks the time of entry against the current time, and batch deletes any which have been in the store for longer than the one-hour window.  
    
The getClosestPeers query is a direct query of the peers taken from the DHT's RoutingTable class, which is responsible for managing the kBuckets. The query looks through all nodes in the kBuckets and returns the closest 20 (as the default bucket size in IPFS is 20). Libp2p uses the javascript implementation k-bucket \footnote{K-bucket: https://github.com/tristanls/k-bucket} to handle the management of the buckets. The function does a raw calculation of the XOR distances by comparing each PeerId in the bucket as a unit8aray to the CID as a uint8array and then orders them from nearest to furthest. 
    
With a populated routing table, it is now possible to query the network to find any provider of the created CID. In this instance, the very first initialised node – who only had contact details for the second initialised node – can query the DHT using the built-in \textit{findProviders} function. The result of the promise is an array containing the details of any node providing the requested content. More details on the system's configuration can be found in Appendix A.
    
\vspace{-16pt}

\section{Evaluation}
\vspace{-32pt}

\begin{figure}[!h]
  \centering
  \makebox[\textwidth][c]{\includegraphics[width=1.0\textwidth]{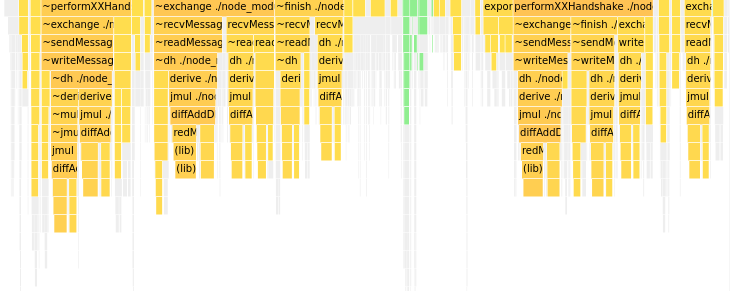}}
  \caption{All processes - With Kad-DHT processes shown in green}
  \label{fig_approach}
\vspace{-26pt}
\end{figure}
To evaluate the runtime performance of the JavaScript implementation of the Kad-DHT, we can examine the flame graph of the running processes. Figure 2 shows the performance of the entire program from start to finish. Each rectangle represents a stack frame, with the y-axis showing the number of frames on the stack – the stack depth. The bottom of each icicle shows the function on-CPU, with everything above it being the function ancestry. The x-axis spans the entirety of the sample population grouped alphabetically. The total width of each rectangle is the total time it was on-CPU or part of the ancestry that was on-CPU; the wider the rectangle, the more CPU consumed per execution. It is worth noting that time is not represented in flame graphs. The Graphs and the logs used to generate them can be found in the corresponding git repo\footnote{Code can be found at: https://github.com/aaoi990/ipfs-kad-dht-evaluation/tree/main/perf}. 
\vspace{-30pt}
\bgroup
\def\arraystretch{1.3} 
\begin{table}
\centering
\caption{CPU time by package based on Fig.~\ref{fig_approach}.}
\setlength{\tabcolsep}{3.2mm} 
\begin{tabular}{|c c c|} 
    \hline
    \textbf{Package} & \textbf{Function} & \textbf{Percentage} \\ 
    \hline 
    libp2p-noise & performXXhandshake & 28.9\\ 
    libp2p-noise & exchange & 18.87\\
    libp2p-noise & finish & 10.07\\
    peer-id & createFromPubKey & 4.86\\
    libp2p & encryptOutbound & 2.43 \\
    libp2p & encryptInbound & 2.005\\
    \hline
\end{tabular}
\vspace{-25pt}
\end{table}
\egroup

Figure 2 illustrates that unsurprisingly the vast majority of CPU usage was spent in the crypto functions, either performing handshakes between nodes or in the functions that support the key generation process.
\vspace{-20pt}

\begin{figure}[!h]
  \centering
  \makebox[\textwidth][c]{\includegraphics[width=1.0\textwidth]{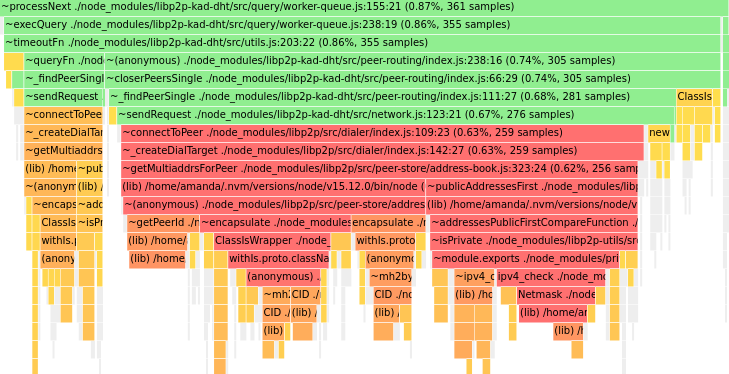}}
  \caption{Some of the Kad-DHT specific processes}
  \label{fig_approach2}
\vspace{-22pt}
\end{figure}

The key generation for a basic Libp2p2 node is a base64 encoded string of a protobuf containing a DER-encoded buffer. A node buffer is then used to pass the base64 protobuf to the multi hash function for the final PeerId generation. By default, the public key is 2048 bit RSA. As suggested in the security improvements in \cite{S/Kad}, peerId generation should be an expensive process in order to mitigate the ease of performing Sybil attacks, and although it was expensive compared to the overall effort of the program, this was primarily because of the default usage of RSA. If EC had been used as per CVE-2020-10937 \cite{eclipse}, the CPU overhead would have been significantly lower. Figure 3 illustrates one of the full stack depths with Kad-DHT ancestry. 

\vspace{-28pt}

\bgroup
\def\arraystretch{1.3} 
\begin{table}
\centering
\caption{CPU time by DHT component based on Fig. 3.}
\setlength{\tabcolsep}{3.2mm} 
\begin{tabular}{|c c c|} 
    \hline
    \textbf{Package} & \textbf{Function} & \textbf{Percentage} \\ 
    \hline 
    network & writeReadMessage & 1.08\\
    worker-queue & processNext & 0.87\\ 
    peer-routing & closerPeersSingle & 0.4 \\
    routing & add & 0.1 \\
    index & nearestPeersToQuery & 0.1 \\
    \hline
\end{tabular}
\vspace{-41pt}
\end{table}
\egroup

\begin{figure}[!h]
  \centering
  \makebox[\textwidth][c]{\includegraphics[width=1.0\textwidth]{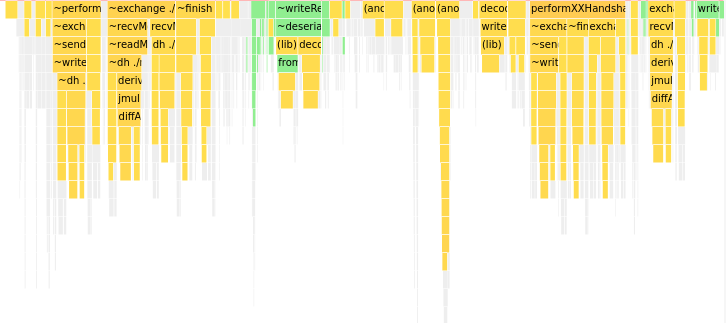}}
  \caption{Kad-DHT processes over an one-hour window}
  \label{fig_approach3}
\vspace{-18pt}

\end{figure}

Overall the Kad-DHT functions occupied a very low percentage of the CPU time, consistently presenting at less than 3.00\%, with the highest usage coming from network functions. The test code being run is a simple start – provide – find – stop sequence, meaning the bulk of the work is being done to configure, connect and route the nodes. It is expected that the longer the program runs, the greater percentage of time the Kad-DHT functions would occupy due to the routing table maintenance functions. During normal operations, the Kad-DHT will force a refresh every 10 minutes by default. During this, each bucket is gone through - from bucket 0 up until the highest bucket that contains a peer (currently capped at 15). A random address from the address space that could fit in the chosen bucket is then selected, and a lookup is done to find the k closest peers to that random address. This constantly ensures that each bucket is filled with as many peers that will fit. Figure 4 results from timing the original code to run for an one-hour window, enabling multiple routing table refreshes. In the timed run, Kad-DHT functions accounted for 11.58\% of CPU usage up from the initial program run of 2.55\%, which is a 354\% increase in the amount of time spent in functions with Kad-DHT ancestry.
 
\vspace{-16pt}

\section{Conclusions}
\vspace{-12pt}
eGovernance presents unique challenges in terms of privacy-preserving and providing secure solutions in eGovernance services. Precisely when the utilized data is derived from industrial control systems and sensors. In this paper, we present GLASS, our decentralized solution, that moves towards that direction by examining the effectiveness and efficiency of distributed cutting-edge technologies and demonstrates the capacity of a public, distributed infrastructure based on the InterPlanetary File System (IPFS). 

One practical implementation of the GLASS concept is being done within the aims of the GLASS project, highlighting how the GLASS concept can potentially be integrated into a broad field of use cases. Our proposed GLASS-oriented approach is a decentralized solution that combines the InterPlanetary File System (IPFS) with Distributed Ledger Technology and Smart Contracts to secure eGovernance services. We show in this paper how our approach can be used to fulfil the needs of the GLASS concept. Finally, and on top of the above, we created a testbed environment to measure the IPFS performance.

\vspace{-12pt}

\section*{Acknowledgments}
\vspace{-8pt}

The research leading to these results has been partially funded by the European Union's Horizon 2020 research and innovation programme, through funding of the GLASS project (under grant agreement No 959879).
\vspace{-16pt}
\bibliographystyle{splncs}
\bibliography{references}
\vspace{-18pt}

\appendix 

\section{Appendices}
\vspace{-8pt}

\subsection{Libp2p node initialisation}
\label{app:a}

\vspace{-8pt}
\begin{lstlisting}[numbers = none, caption = {Libp2p node initialisation.}, basicstyle=\small]
    
    const node = await Libp2p.create({
    addresses: {
      listen: ['/ip4/0.0.0.0/tcp/0']
    },
    modules: {
      transport: [TCP],
      streamMuxer: [Mplex],
      connEncryption: [NOISE],
      dht: KadDHT,
    },
    config: {
      dht: {
        kBucketSize: 20,
        enabled: true,
        randomWalk: {
          enabled: true,
          interval: 300e3,
          timeout: 10e3
        }
      }
    }
  })
    \end{lstlisting}
\vspace{-8pt}

\subsection{Random walk PeerId creation}
\label{app:b}
\vspace{-8pt}

\begin{lstlisting}[numbers = none, caption = {Random walk PeerId creation.}, basicstyle=\small]
    
    const digest = await multihashing(
        crypto.randomBytes(16), 'sha2-256')
    const id = new PeerId(digest)
    \end{lstlisting}
\vspace{-20pt}

\subsection{Transforming content to a CID}
\label{app:c}
\vspace{-8pt}
 
 \begin{lstlisting}[numbers = none, caption = {Transforming content to a CID.}, basicstyle=\small]
    
    const hash = crypto.createHash('sha256')
        .update('hello world!').digest()
    const encoded = multihash.encode(hash, 'sha2-256')
    const cid = new CID(multihash.toB58String(encoded))
    \end{lstlisting}
\vspace{-20pt}

\subsection{A node providing content}
\label{app:d}
\vspace{-8pt}

\begin{lstlisting}[numbers = none, caption = {A node providing content.}, basicstyle=\small]
    
    await node.contentRouting.provide(cid)
    \end{lstlisting}
\vspace{-20pt}

\subsection{Distributing content to the closest peers}
\label{app:e}
\vspace{-8pt}

\begin{lstlisting}[numbers = none, caption = {Distributing content to the closest peers.}, basicstyle=\small]
    
    async provide (key) {
    dht._log(`provide: ${key}`)

      /** @type {Error[]} */
      const errors = []

      // Add peer as provider
      console.log('starting to provide')
      await dht.providers.addProvider(key, dht.peerId)

      const multiaddrs = dht.libp2p ? dht.libp2p.multiaddrs : []
      const msg = new Message(Message.TYPES.ADD_PROVIDER, key.bytes, 0)
      msg.providerPeers = [{
        id: dht.peerId,
        multiaddrs
      }]

      async function mapPeer (peer) {
        dht._log(`putProvider ${key} to ${peer.toB58String()}`)
        try {
          await dht.network.sendMessage(peer, msg)
        } catch (err) {
          errors.push(err)
        }
      }

      // Notify closest peers
      await utils.mapParallel(dht.getClosestPeers(key.bytes), mapPeer)

      if (errors.length) {
        throw errcode(new Error(`Failed to provide to ${errors.length} of ${dht.kBucketSize} peers`), 'ERR_SOME_PROVIDES_FAILED', { errors })
      }
    },
    
    \end{lstlisting}
  \vspace{-20pt}
  
\subsection{Creation of the datastore}
\label{app:f}
\vspace{-8pt}
 
  \begin{lstlisting}[numbers = none, caption = {Creation of the datastore.}, basicstyle=\small]
    
    const dsKey = [ 
        makeProviderKey(cid),'/',
        utils.encodeBase32(peer.id)].join('') 
    const key = new Key(dsKey) 
    const buffer = Uint8Array.from(
        varint.encode(time.getTime())) 
    store.put(key, buffer) 
    \end{lstlisting}
\vspace{-20pt}

\subsection{Calculating the closest Peers using the XOR metric}
\label{app:g}
\vspace{-8pt}

     \begin{lstlisting}[numbers = none, caption = {Calculating the closest Peers using the XOR metric.}, basicstyle=\small]
    
    closest (id, n = Infinity) {
    ensureInt8('id', id)
    
    if ((!Number.isInteger(n) && n !== Infinity) || n <= 0) {
      throw new TypeError('n is not positive number')
    }
    let contacts = []

    for (let nodes = [this.root], 
        bitIndex = 0; nodes.length > 0 && contacts.length < n;) {
      
      const node = nodes.pop()
      if (node.contacts === null) {
        const detNode = this._determineNode(
            node, id, bitIndex++)
        nodes.push(
            node.left === detNode ? node.right : node.left)
        nodes.push(detNode)
      } else {
        contacts = contacts.concat(node.contacts)
      }
    }

    return contacts
      .map(a => [this.distance(a.id, id), a])
      .sort((a, b) => a[0] - b[0])
      .slice(0, n)
      .map(a => a[1])
  }
    \end{lstlisting}
 \vspace{-20pt}
 
\subsection{Finding providers}
\label{app:h}
\vspace{-8pt}
  
   \begin{lstlisting}[numbers = none, caption = {Finding providers.}, basicstyle=\small]
   
    await all(nodes[0].contentRouting
        .findProviders(cid)) 
   \end{lstlisting}
 \vspace{-20pt}
  
   

\subsection{Result of the ``Finding Providers'' query}
\label{app:i}
\vspace{-8pt}
    \begin{lstlisting}[numbers = none, caption = {Result of the findProviders query.}, basicstyle=\small]
    
   {
    id: PeerId {
      _id: <Buffer 12 20 83 42 f7 0e 33 90 d1 c4 41 d0 80 d7 16 63 be 43 95 20 3c b1 79 5e 23 d7 28 12 3e 4a 0f aa d9 d3>,
      _idB58String: 'QmXB3LoMkXQh3HzQo1fy-
        9UEJZZQw2MmJKWRhG4nfbTR7Qe',
      _privKey: undefined,
      _pubKey: undefined
    },
    multiaddrs: []
  }
  \end{lstlisting}
    
\end{document}